\documentstyle[psfig,graphics]{aa}

\def\ros{{\sl ROSAT}}
\def\etal{{et\,al.}}
\def\ergsec{\hbox{erg s$^{-1}$}}
\def\ergcm{\hbox{erg cm$^{-2}$ s$^{-1}$}}
\def\msun{M$_{\odot}$}

\def\it{\sl}
\def\degs{\ifmmode ^{\circ}\else$^{\circ}$\fi}
\def\amin{\ifmmode ^{\prime}\else$^{\prime}$\fi}
\def\asec{\ifmmode ^{\prime\prime}\else$^{\prime\prime}$\fi}
\def\fss{\hbox{$.\!\!^{\rm s}$}}        
\def\farcs{\hbox{$.\!\!^{\prime\prime}$}}  
\def\h{$^{\rm h}$}\def\m{$^{\rm m}$}
\def\s{$^{\rm s}$}
\newbox\grsign \setbox\grsign=\hbox{$>$}
\newdimen\grdimen \grdimen=\ht\grsign
\newbox\laxbox \newbox\gaxbox
\setbox\gaxbox=\hbox{\raise.5ex\hbox{$>$}\llap
     {\lower.5ex\hbox{$\sim$}}}\ht1=\grdimen\dp1=0pt
\setbox\laxbox=\hbox{\raise.5ex\hbox{$<$}\llap
     {\lower.5ex\hbox{$\sim$}}}\ht2=\grdimen\dp2=0pt
\def\gax{\mathrel{\copy\gaxbox}}
\def\lax{\mathrel{\copy\laxbox}}
\def\s13{RX\,J0045.5+4206}
\def\dashdot{\cdots\cdots\cdots\cdots\cdots\cdots\cdots\cdots\cdots\cdots\cdots\cdots\cdots\cdots\cdots\cdots\cdots\cdots}

\begin{document}

\voffset=15mm

   \thesaurus{06         
              (08.09.2;  
               08.23.2;  
               09.09.1;  
               09.18.1;  
               11.09.1   
                      )} %

  \title{The WC6 Wolf-Rayet star MLA\,1159 in M\,31 and its ionization nebula 
    BA\,1-642}

  \author{J. Greiner\inst{1} \and G. Tovmassian\inst{2} \and 
          S. Komossa\inst{3} \and M. Rosado\inst{2} \and A. Arrieta\inst{2}}

  \offprints{J. Greiner, jgreiner@aip.de}

   \institute{Astrophysical Institute
        Potsdam, An der Sternwarte 16, D-14482 Potsdam, Germany
       \and
        Instituto de Astronom\'{\i}a, UNAM, Apdo. Postal 877, 22800 Ensenada, 
         B.C., M\'{e}xico
       \and
        Max-Planck-Institute for extraterrestrial Physics,
        D-85740 Garching, Germany
       }

   \date{Received 22 December 1998 / Accepted 14 April 1999}

   \titlerunning{WC6 star with ionization nebula in M31}

\maketitle

\begin{abstract}
We report on optical imaging and spectroscopic observations of the Wolf-Rayet
candidate star MLA 1159 and the surrounding ionization nebula BA\,1-642
in the Andromeda Galaxy. Though both objects have been known for many years, 
our observations
(1) confirm the Wolf-Rayet nature of MLA 1159,
(2) allow one to determine the nebula as an ionization nebula, and
(3) demonstrate the association of MLA 1159 with the nebula. 
The supersoft X-ray source
\s13\ whose error box encompasses the full nebula, seems to be a chance 
superposition and not related to MLA 1159 and/or BA 1-642.

\end{abstract}

\section{Introduction}
Supersoft X-ray binaries,
highly luminous
($L_{bol} \sim 10^{36}-10^{38}$ ergs s$^{-1}$) and low-temperature (20--40 eV) 
X-ray sources, were established as a new class of objects by ROSAT
(Tr\"umper \etal\ 1991, Greiner \etal\ 1991, van den Heuvel \etal\ 1992). 
The archetype supersoft binary,
CAL 83, exhibits a 25 pc ionization nebula (Pakull \etal\ 1989,
Remillard \etal\ 1995).
Aimed at identifying supersoft X-ray sources in the Andromeda Galaxy via
an ionization nebula we have searched for
extended optical emission within the X-ray error boxes of the 15 
supersoft sources within M\,31 (Greiner \etal\ 1996). One optical
nebula was found (Fig. \ref{fc}), and we obtained detailed optical data -- 
spectroscopy as well as narrow-band imaging -- to determine its relation
to the positionally correlated supersoft X-ray source RX J0045.5+4206.

Anticipating the identification of the central star as a Wolf-Rayet star,
we will briefly introduce this stellar type and its relation to nebulae before
proceeding with the observations.
Wolf-Rayet (WR) stars were first identified by their broad emission lines
(Wolf \& Rayet 1867). These broad emission lines are thought to be
indicative of a thick expanding stellar atmosphere. The line widths
imply terminal velocities of 1000--3000 km/s, suggesting that the
outflowing stellar material is gravitationally unbound. WR stars have
typical mass loss rates of a few times 10$^{-5}$ \msun/yr and are the 
most powerful sources of stellar winds among massive early-type
stars (Barlow 1982, Abbott \& Conti 1987, Maeder \& Conti 1994).

A considerable fraction of WR stars, though not all, are surrounded by 
ring nebula (e.g. Miller \& Chu 1993, Marston 1997) which are thought to 
be leftovers of the evolutionary phase when
the massive WR-progenitors stripped off their outer envelopes.
Theoretically one would expect two shells around a WR star according to
the evolutionary stages, i.e. a large fossil bubble of ISM swept up during the
main-sequence stage, and a small bubble of circumstellar material blown
away from the WR wind. Follow-up studies of the WR nebulae have shown,
however, that not all ring nebulae are wind-blown bubbles consisting of ISM
(Weaver \etal 1977).
In some cases the nebulae contain mostly stellar matter with which the
WR wind is interacting, while in other cases the nebulae are ionized by
the UV flux of the central WR star.

The population of Wolf-Rayet stars in the Milky Way and the Local Group has 
been the subject of ample studies (e.g. Massey 1998, Massey \& Johnson 1998). 
It is important not only for the study of each 
particular star, but also because of their role in stellar evolution as well as
their contribution to and interaction with the interstellar matter. 
It provides also evolutionary information about the galaxies they belong to. 
There have been systematic searches of WR stars in our Galaxy, the 
Magellanic Clouds, M\,31 and M\,33 as well as several other nearby galaxies
(e.g. Massey 1999, Massey \& Johnson 1998).

M 31 has been surveyed several times with different goals. In an early attempt
to establish the existence of gaseous nebulae, Baade \& Arp (1964) compiled
a list of 688 emission nebula. One of these nebulae, called BA 1-642,
is the nebula which we found during our cross-correlation with supersoft
X-ray sources. The same object is also listed as PAV78 915 in the catalog
of Pellet \etal\ (1978). 
A systematic search for WR stars in M\,31 was conducted largely by 
Moffat \&~Shara (1983, 1987). 
The central object of this nebula has been classified
as a WR star candidate (object MLA 1159; Meyssonier \etal\ 1993)
based on an objective-prism
survey of M 31 with a dispersion of 2000 \AA/mm in the 4350--5300 \AA\ range.

Here we report the confirmation of this WR star classification and present 
results of a photometric and spectroscopic study of the WR star  
as well as the ring-like nebulae surrounding it.

\section{Observations and Results}

\begin{figure}[t]
      \vbox{\psfig{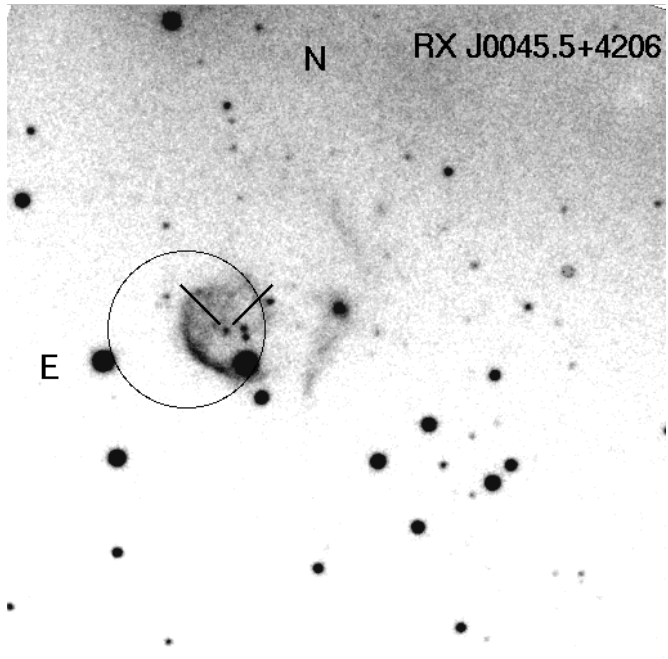}}\par
      \caption[fchart]{A 3\farcm3$\times$3\farcm3 H$\alpha$ image of the 
          field around the Wolf-Rayet star MLA 1159 (marked by two dashes) 
          and the clearly visible nebula.
         The position of the X-ray source RX\,J0045.5+4206 with its 
         3$\sigma$ error of  24\arcsec\, is shown as the overlayed circle.
         Based on the observed optical and X-ray properties, this ROSAT source
         seems not to be associated with the Wolf-Rayet star or the nebula. }
      \label{fc}
\end{figure}

Imaging observations of the region of the X-ray error box of RX J0045.4+4206
(see Fig. \ref{fc})
were performed during three nights in October 19--21, 1996
with the 1.5\,m telescope of the Observatorio Astron\'omico Nacional (OAN) 
de San Pedro M\'artir (SPM), M\'exico, equipped with a 1024$\times$1024
pixel CCD TEK. The total field of view was 4\farcm3$\times$4\farcm3. The 
seeing was typically 1\farcs5, thus we observed with the CCD chip binned to 
2$\times$2 and with a corresponding  pixel scale of 0\farcs51/pixel.
Images were obtained in various filter bands (see Table \ref{filt}, listing
central wavelength, FWHM and peak transmission)
to allow a diagnostic of the excitation level.
The data reduction used standard MIDAS procedures for bias and cosmic ray
subtraction and flat-fielding.

\begin{table}[ht]
\caption{Details on narrow-band filters used}
\vspace{-0.2cm}
\begin{tabular}{cccc}
      \noalign{\smallskip}
      \hline
      \noalign{\smallskip}
   Name & central wave-   & mean        & transmission at \\
        & length (\AA)   & width (\AA) & central w. (\%) \\
 \noalign{\smallskip}
 \hline
 \noalign{\smallskip}
  \ion{He}{ii}     & 4689 & 61   & 59 \\
  cont1            & 4772 & 44   & 50 \\
  H$\beta$(narrow) & 4861 & ~\,7 & NA \\
  H$\beta$(wide)   & 4871 & 49   & 70 \\
  \ion{[O}{iii]}     & 5016 & 51   & 69 \\
  cont2            & 5050 & 50   & NA \\
  \ion{He}{i}      & 5881 & 32   & 62 \\
  H$\alpha$(narrow) & 6564 & 11  & 66 \\
  H$\alpha$(wide)   & 6564 & 72  & 85 \\
  cont3            & 6650 & 47   & 64 \\
  \ion{[N}{ii]}      & 6587 & 11   & 62 \\
  \ion{[S}{ii]}      & 6729 & 52   & 60 \\
      \noalign{\smallskip}
      \hline
    \end{tabular}
   \label{filt}
   \end{table}

Spectroscopy was performed on two occasions in August 1996 and August 1997,
respectively, using the 2.1\,m telescope of OAN SPM.
The Boller \& Chivens spectrograph with a 600 l/mm (1996) and 400 l/mm (1997)
grating was employed to get spectra with overall FWHM resolution of 4.5 \AA\
and 6--7 \AA. The seeing was typically 1\farcs5, and consequently
a 2\asec\ slit has been used. The data reduction again used standard MIDAS 
procedures of the long-slit spectrum reduction package.

Finally, we also observed the nebula with the UNAM Fabry-Perot interferometer 
PUMA. The Fabry-Perot observations of the  nebula BA 1-642
were carried out during the night of November 2, 1997
at the f/7.5 Cassegrain focus of the 2.1~m telescope of the OAN SPM using the
UNAM Scanning Fabry--Perot Interferometer PUMA (Rosado \etal\ 1995).

A 1024$\times$1024 thinned Tektronix CCD detector, with
an image scale of 0.59 arcsec~pixel$^{-1}$, was used with a 2$\times$2
on-chip binning in both dimensions.  Thus, the resulting image format was 
512$\times$512 pixels, with a spatial resolution of 1.18 arcsec~pixel $^{-1}$.

An interference filter centered at $\lambda = 6570$ \AA\ and having a
bandpass of 20 \AA\ was used in order to isolate the H$\alpha$ line.
The scanning Fabry-Perot interferometer is an ET-50 of Queensgate
Instruments  with a servo-stabilization system. The main characteristics
of this interferometer are: interference order of 330, free spectral range of
18.92 \AA\ (equivalent to a velocity range of 908
km~s$^{-1}$) and sampling spectral resolution of 0.41 \AA\
(equivalent to 18.9 km~s$^{-1}$) at the wavelength of the H$\alpha$ line, 
achieved by scanning the interferometer gap at 48 positions.
Thus, the resulting data cubes have dimensions of 512$\times$512$\times$48.

Under these conditions, we have obtained one nebular data cube with an
exposure time of 48 min.  We also obtained  two calibration data cubes spaced 
at the beginning and at the end of the observations in order to check for 
possible flexures of the equipment. For the calibration cubes we have used
the Ne line at $\lambda$ 6598.95 \AA.

\begin{figure*}
  \resizebox{12.cm}{!}{\includegraphics{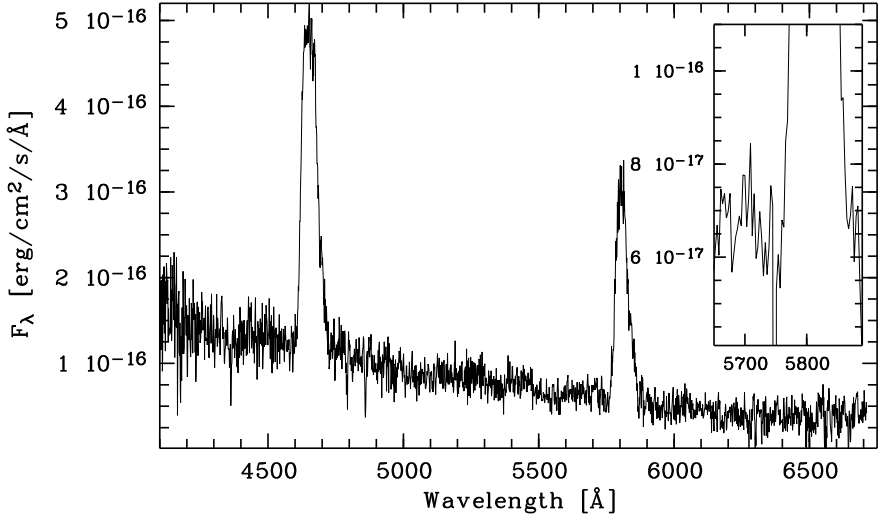}}
  \hfill
  \parbox[b]{55mm}{
  \caption{Mean spectrum of the star MLA 1159. The broad, strong emission 
    lines of \ion{C}{III}, \ion{C}{IV} 4650 \AA\ and 
    \ion{C}{IV} 5805 \AA\ clearly identify this as a C-type Wolf-Rayet star.
     Weak \ion{C}{IV} 4442 \AA\ and \ion{C}{III} 5696 \AA\ (see inset) 
     emission is also visible.}
   }
  \label{wrspec}
\end{figure*}

The data reduction and analysis were performed using the specific reduction
package CIGALE (Le Coarer \etal\ 1993). This software was used to remove
the cosmic rays, to carry out the wavelength calibration
of the data cubes, to obtain continuum subtracted $\lambda$-cubes 
and to carry out the emission line profile analysis.

\subsection{The central star MLA 1159}

The spectrum of the central star, MLA 1159 (Fig. \ref{wrspec}), shows strong 
and broad emission 
lines at 4650 \AA\ and 5805 \AA\ (which we interpret as \ion{C}{III}
and \ion{C}{IV}), and clearly suggest a carbon Wolf-Rayet star
thus confirming the Wolf-Rayet candidacy proposed by Meyssonier \etal\ (1993).
The ratios of \ion{C}{III} 5696 / \ion{C}{IV} 5805 $\ll$ 1 and
\ion{C}{III} 5696 / \ion{O}{V} 5592 $>$ 1 identify MLA 1159 without
much doubt as WC6 subtype (according to Smith 1968, van der Hucht 1981).
The recently proposed refinement of the WC classification scheme 
(Smith \etal 1990, further developed by Crowther \etal\ 1998) does not alter 
this result.
With a FWHM of the \ion{C}{IV} 5805 \AA\ line of $54 \pm 5$ \AA\ and
$\log (\ion{C}{IV} 5805 / \ion{C}{III} 5696) = 1.0 \pm 0.2$, we again 
arrive at a WC6 subtype (see Table 3 in Crowther \etal\ 1998).

With a measured $m_{\rm V}$ = 19.2 mag, using a distance of 650 kpc
and assuming only galactic foreground absorption and no intrinsic M\,31
extinction, i.e. A$_{\rm V}^{\rm tot}$=0.6 mag,
we derive $M_{\rm V}$ = --5.7$\pm$0.3 mag, at the bright end of the range of 
absolute magnitudes of WC5-6 stars of $M_{\rm V}$ = --3.9$\pm$0.5 mag
(Lundstr\"om \& Stenholm 1984).

We derive a position based on the digitized sky survey of 
RA. (2000) = 00\h 45\m 31\fss0, Decl. (2000) = 42\degr 06\amin57\asec\
(error of $\pm$1\asec) which is consistent with the position of MLA 1159
(Meyssonier \etal\ 1993).

\subsection{The nebula BA 1-642 $\equiv$ PAV78 915}

\subsubsection{F--P interferometry}

\begin{figure*}
  \resizebox{11.78cm}{!}{\includegraphics{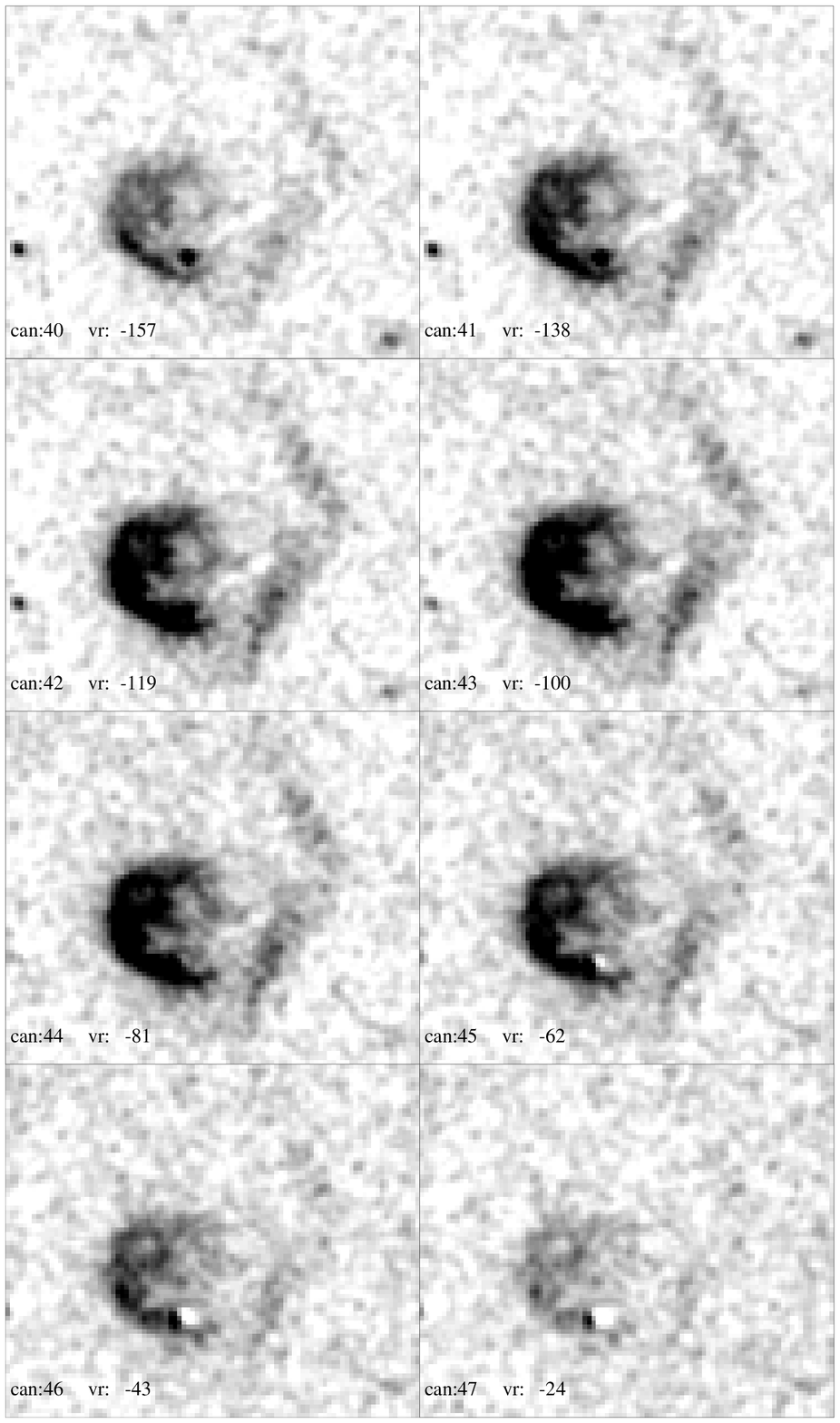}}
  \caption{1\farcm8$\times$1\farcm5 H$\alpha$ images ($\lambda$-maps) of the 
      nebula obtained in 
      different velocity channels with the F--P interferometer PUMA.
      The other PUMA channels (0--39) do not show appreciable nebular emission.
      The channel numbers and heliocentric velocities are marked at the lower 
      left corners of each image. The images are continuum subtracted.}
  \label{fpall}
\end{figure*}

Fig. \ref{fpall} shows the PUMA $\lambda$-maps corresponding to the velocity 
channels where the nebula BA 1-642 is detected. We see that it covers a 
velocity range from V$_{helio}$ = --157 km~s$^{-1}$ to --24 km~s$^{-1}$ 
having maximum intensity at   V$_{helio} \approx$  --100 km~s$^{-1}$. 
It is interesting to note that other M31 
nebulae are also detected in the 10 arcmin field of the PUMA at
the same velocities but with a narrower range in velocity. This suggests that
BA 1-642 has internal motions larger than the internal motions of classical
HII regions.

\begin{figure}
  \resizebox{8.7cm}{!}{\includegraphics{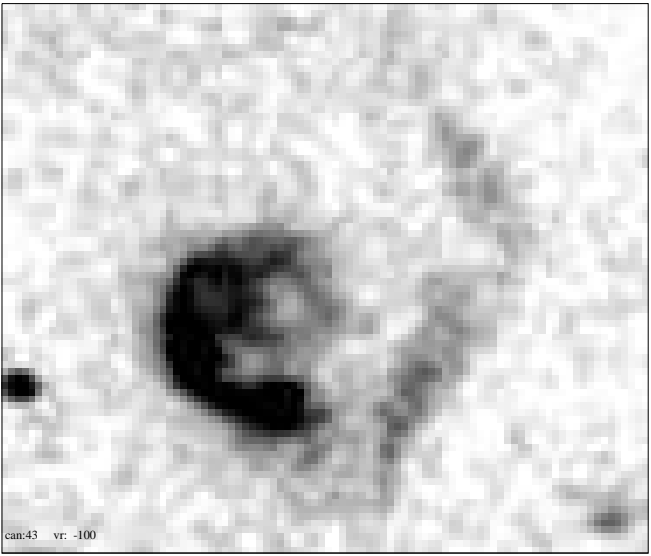}}
  \caption{The enlarged image of the nebular complex in the H$\alpha$ filter, 
    in the velocity channel 43, corresponding to $V_{\it helio}=-100$ km/sec.
    Note the large circular shell centered approximately on the northern 
    edge of the compact bubble. }
  \label{fpha}
\end{figure}

Fig.  \ref{fpha} is an enlargement of one of the velocity channels (at 
 V$_{helio}$ = --100 km~s$^{-1}$). 
The WR star MLA 1159 is located within a nearly circular nebula
(see also Figs. \ref{fc}, \ref{nebima}). The south-east rim of the nebula is 
very pronounced, particularly in the H$\alpha$ line.

However, from this figure we also note, that there is some extended 
structure on the West and the total complex probably consists of  two 
different sets of nebulosities: a nearly hemispherical internal bubble 
with an angular size of 34$\times$26 arcsec corresponding to a linear size of 
107$\times$82 pc at the adopted distance of 650 kpc to M31 (this is 
BA 1-642 $\equiv$ PAV78 915) and a V-shaped nebulosity to the West of
the internal bubble (note that this part has its own nomenclature, i.e.
BA 1-641 $\equiv$ PAV78 913). A close inspection of Fig. \ref{fpha} suggests 
that this second
nebulosity seems to be a thick shell of 74 arcsec in diameter (equivalent 
to 233 pc) centered at the Northern edge of the internal bubble. The external 
shell 
is brighter near the MLA 1159  which is probably photoionizing this nebulosity.
The fact that both nebulosities have their maximum intensities in the same 
velocity channel implies that they would be at the same galactocentric 
distance and consequently a physical link between them is favored against a 
chance superposition along the line of sight. The thick shell could be a 
trace of the photoionization or the interaction with the interstellar medium
of the winds 
of the   progenitor of the WR star or other massive stars. Deeper observations 
are required in order to confirm this hypothesis. In the following we will 
restrict our kinematical study to the internal bubble.

We have obtained radial velocity profiles integrated in several zones of 
the bubble, the brightest region of the shell and other HII regions of the 
field. We confirm that the shell has the same velocity as the borders of the 
bubble: V$_{helio} \approx$  --100 km~s$^{-1}$ . Even if  our velocity 
profiles have 
low S/N ratio (about 4) we are able to detect two heliocentric velocity 
components for the central region of the bubble: at --83 and --131 km~s$^{-1}$
with an uncertainty of about 4 km~s$^{-1}$. 
On the other hand, the velocity profiles at the edges of the bubble are simple
and can be fitted by a single Gaussian function. This could be interpreted as 
an expansion motion with a velocity of 25  km~s$^{-1}$. Consequently, we can 
estimate the kinematic age of the nebula using t$_{kin}$ = 0.6 R/V
where t$_{kin}$ is in units of 10$^6$ yr, R in pc and V in km~s$^{-1}$
(taken from Weaver \etal\ 1977 models for the evolution of wind blown
bubbles, see also Rosado 1986). 
Taking R = 54 pc and V = 25  km~s$^{-1}$, we obtain a kinematic age of 
2.1 10$^{6}$ yr. This value of the kinematic age is larger than the duration
of the WR phase. Thus, this bubble could not be 
formed by the wind of MLA 1159 (i.e. W-type, Chu 1981, Rosado 1986). 
Instead, this bubble should be formed by the WR star progenitor and is only 
illuminated by the WR star
(i.e. R$_{s}$-type, Chu 1981, Rosado 1986). This is typical 
for the nebulae associated with WC-type stars.

\subsubsection{Narrow-Band Photometry and Spectrophotometry}

The shape and the size of the
nebula is very similar at different wavelengths (see Fig. \ref{nebima} with
images of the nebula in the six bands \ion{He}{II}, H$\beta$, \ion{[O}{III]},
H$\alpha$, \ion{[N}{II]} and \ion{[S}{II]}).
The nebula shows strong Balmer emission lines, \ion{[O}{II]} and \ion{[S}{II]} 
but weak \ion{[O}{III]} (see Fig. 6).
We note that the emission lines are all blue-shifted by 3--4 \AA, consistent
with the above reported  velocities.

It is not completely clear whether the ring-like nebula is physically
center filled or not, because it shows less emission in the inner parts.
Nevertheless, we have measured the size for both the inner and the outer
ring based on the pronounced, bright south-east rim visible in the 
H$\alpha$ image to $R_{\rm in} \sim 11$\asec\
and $R_{\rm out} \sim 14$\asec, respectively. At the distance of M\,31 this 
corresponds to radii of 34 and 50 pc, respectively.

Using the ratio of \ion{[O}{III]} 5007 \AA/H$\beta = 0.45 \pm 0.08$ as 
criterion, the excitation class is estimated to be 0.5 according to the 
system of Feast (1968) and
Webster (1975). This and the lack of nebular \ion{He}{II} emission lines
indicate that  BA 1-642 is a low-excitation nebula.

The integrated H$\beta$ line flux in regions Ib and IIb is
3$\times$10$^{-15}$ and 2$\times$10$^{-15}$ \ergcm. The analysis of the 
H$\beta$ filter image shows that region Ib corresponds to the mean 
H$\beta$ flux integrated over the full nebula. We therefore scaled the
integrated H$\beta$ line flux from region Ib by the fractional area of
the extraction region to derive a total H$\beta$ line luminosity of the
nebula of 4$\times$10$^{36}$ erg/s.
This H$\beta$  line luminosity is a few percent of the
luminosity of the Wolf-Rayet star, and thus can be produced by the irradiation
of the central Wolf-Rayet star.


\begin{table}
 \caption[]{Line intensity ratios relative to H$\beta$. The values for regions
  a and b are the mean of the sectors I and II, while the values for region 
  III is the mean of IIIa and IIIb (see Fig. \ref{nebima} for the relative
  locations of the different regions). Only the spectra taken of region III
  cover the sulphur lines.}
  \vspace{-0.2cm}
  \begin{tabular}{rcccc}
      \noalign{\smallskip}
      \hline
      \noalign{\smallskip}
   Line~~ & Region a & Region b & Region Ic & Region III \\
 \noalign{\smallskip}
 \hline
 \noalign{\smallskip}
  $\!\!\!$\ion{[O}{III]} 4959 & 0.12$\pm$0.03 & 0.19$\pm$0.06 & 0.18$\pm$0.05 
      & 0.18$\pm$0.02 \\
  $\!\!\!$\ion{[O}{III]} 5007 & 0.38$\pm$0.03 & 0.39$\pm$0.04 & 0.54$\pm$0.06 
      & 0.46$\pm$0.03 \\
  He I 5875           & 0.13$\pm$0.02 & 0.14$\pm$0.02 & 0.12$\pm$0.03 
      & 0.17$\pm$0.02 \\
  \ion{[O}{I]} 6300   & $<$0.01       & $<$0.03 & $<$0.03 
      & 0.06$\pm$0.03 \\
  \ion{[N}{II]} 6548  & 0.33$\pm$0.02 & 0.32$\pm$0.02 & 0.18$\pm$0.05 
      & 0.33$\pm$0.15 \\
  \ion{[N}{II]} 6584  & 1.04$\pm$0.02 & 1.02$\pm$0.03 & 0.78$\pm$0.06 
      & 0.87$\pm$0.20 \\
  \ion{[S}{II]} 6716  & -- & -- & -- 
      & 0.45$\pm$0.06 \\
  \ion{[S}{II]} 6731  & -- & -- & -- 
      & 0.32$\pm$0.05  \\
      \noalign{\smallskip}
      \hline
    \end{tabular}
   \label{ratio}
   \end{table}

\subsubsection{Photoionization modelling}

\begin{figure*}
   \vbox{\psfig{figure=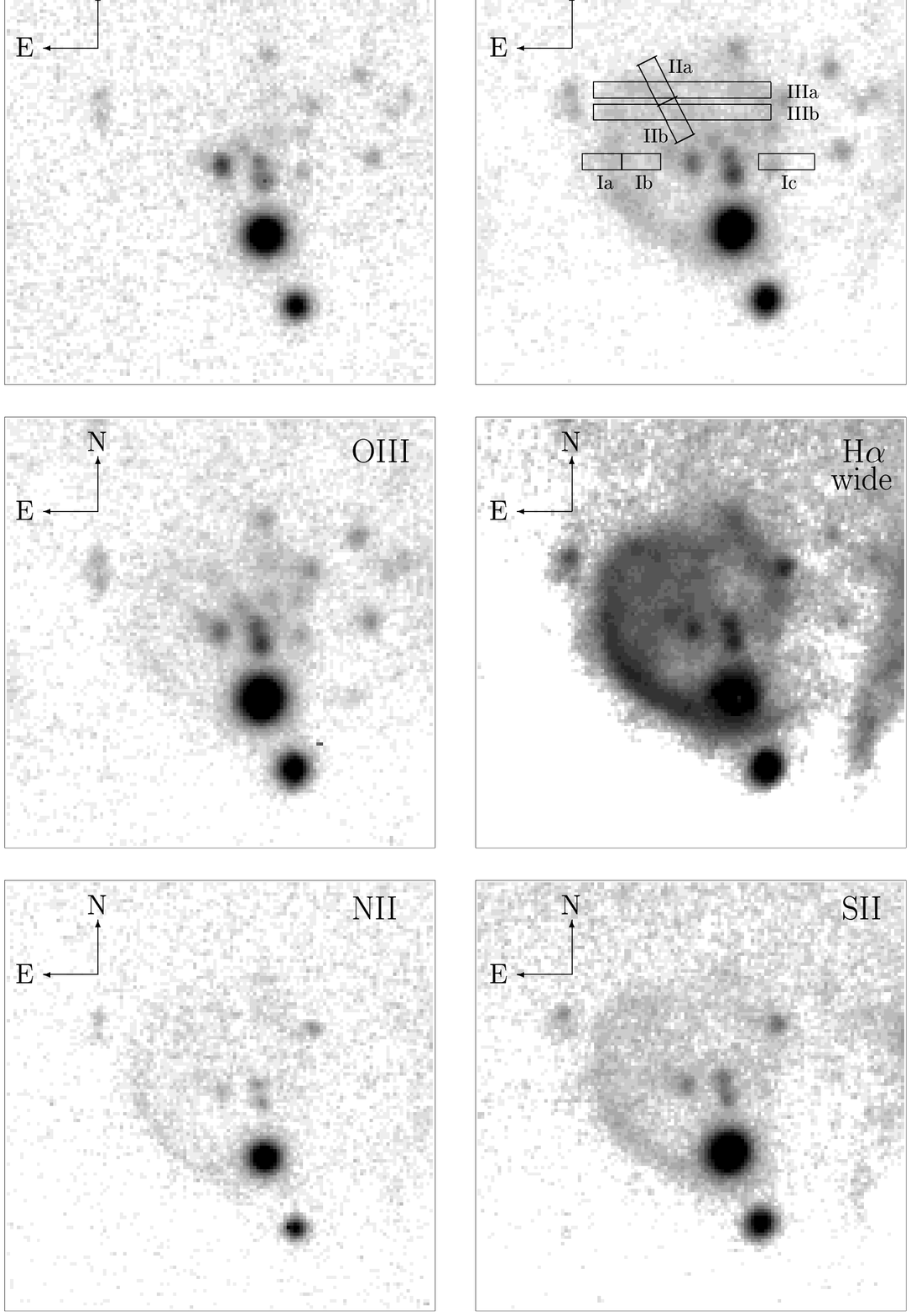,width=14.cm,%
         bbllx=2.cm,bblly=4.1cm,bburx=18.6cm,bbury=28.9cm,clip=}}\par
    \caption[neb]{The nebula BA 1-642 in 6 different filter bands:
         \ion{He}{ii} (top left; 1200 sec exposure time), 
       H$\beta$ (top right; 2400 sec), 
      \ion{[O}{iii]} (middle left; 2400 sec), 
      H$\alpha$ wide (middle right; 1800 sec),
      \ion{[N}{ii]} (bottom left; 1200 sec), 
      \ion{[S}{ii]} (bottom right; 1200 sec).
      All images are centered on the Wolf-Rayet star which is particularly 
      bright in the \ion{He}{II} image due to the strong carbon emission
      (see Fig. \ref{wrspec}).
      Images with exposure times larger than 1200 sec are sums of two
      individual exposures. The slit positions and extraction areas of 
      the long-slit spectroscopy are drawn on top of the H$\beta$ image.
          }
    \label{nebima}
\end{figure*}

In the following, we derive some constraints from the emission line ratios
measured at various locations of the nebula (see Table \ref{ratio} and 
Fig. \ref{nebima}).
We note that all measured Balmer decrements are consistent with the 
recombination value or only very small reddening, certainly lower than
the global total galactic column density in this direction of
$N_{\rm H}^{\rm gal} = 8\times10^{20}$ cm$^{-2}$ (Dickey \& Lockman 1990) 
which corresponds to $A_{\rm V} = 0.6$ mag. We therefore have not applied 
a reddening correction.

\begin{figure*}
  \resizebox{12.cm}{!}{\includegraphics{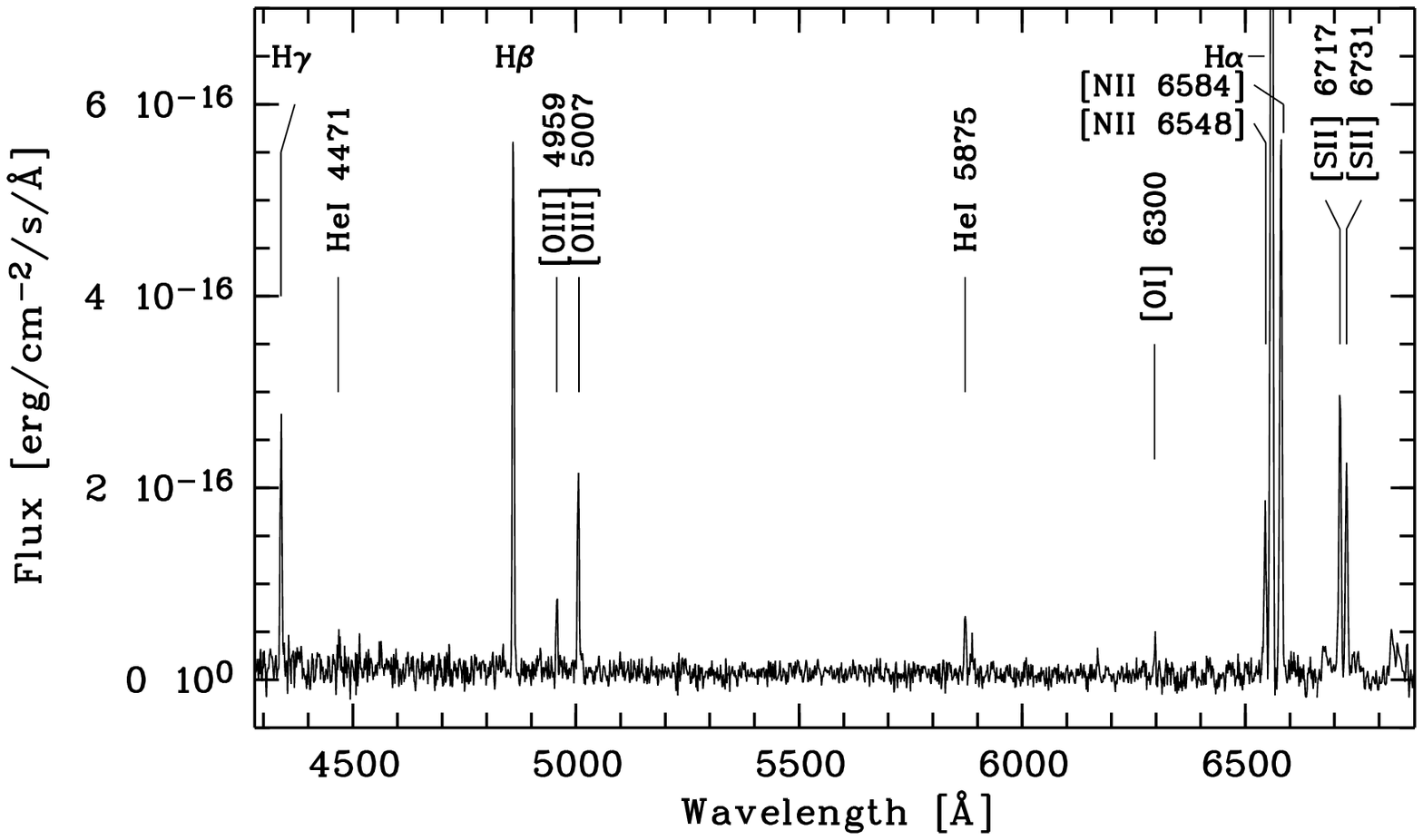}}
  \hfill
  \parbox[b]{55mm}{
  \caption{Mean spectrum of the nebula BA 1-642. Note that H$\alpha$ is 
      cut at the top because the ordinate has been expanded to show also the 
      weak emission lines.}}
  \label{specneb}
\end{figure*}

First, we estimated some of the parameters using diagnostic
emission line ratios:

\noindent {\em Density via sulphur ratio}:
The observed ratio of the two sulphur lines can be used to estimate
the density of the emission-line gas. 
The measured intensity ratio \ion{[S}{II]}$\lambda$6716/6731 = 1.4 is close
to the low-density limit and implies a density 
of  $n \approx 5 \times 10^1$ cm$^{-3}$ (Osterbrock 1989), a value
typical for HII regions.
In the photoionization calculations described below, the gas density
was fixed to this value or varied within a small range of this value 
(within a factor $\sim$2).

\noindent {\em Temperature via oxygen ratio}:
For the inferred gas density, the oxygen intensity ratio 
\ion{[O}{III]}$\lambda$5007/4363 is a useful estimator of the gas temperature.
Given that we only have an upper limit for \ion{[O}{III]}$\lambda$4363,
we derive a corresponding upper limit for the gas temperature of
$T < 19000$ K (Osterbrock 1989). We note that this is
not very strict, but again consistent with what is known for HII regions.  

Second, we have performed photoionization calculations with the code {\em Cloudy}
(Ferland 1993). We assumed the gas clouds to be dust-free (except if 
stated otherwise) and illuminated by the 
continuum of the central Wolf-Rayet star. The continuum shape was 
approximated by a blackbody,
its temperature left free to vary (since there is no one-to-one match in
temperature of a blackbody and a Wolf-Rayet atmosphere). 
Solar gas abundances (Grevesse \& Anders 1989)
were adopted if not stated otherwise. The gas was assumed to be 
ionization bounded. 
The aim of these photoionization calculations is to reproduce the observed 
line-ratios, get clues on the ionizing 
continuum shape (within the limits of the blackbody assumption),
decide whether we see a filled volume or shell, derive various properties of  
the nebula (like ionization parameter, temperature, thickness), 
constrain abundances, and check for the presence of dust.

We have calculated a large grid of photoionization models varying
the ionizing continuum (i.e., $T_{\rm bb}$) and the radius (equivalent to
the  ionization parameter).
Test calculations were also performed for different density, metallicity
and the presence of dust. The ionization parameter is defined as 
\begin {equation}
U=Q/(4\pi{r}^{2}n_{\rm H}c)
\end {equation}
where $Q$ is the number rate of photons above 13.6 eV, $r$ is the distance 
between the Wolf-Rayet star and the emission-line gas, $n_{\rm H}$ is the 
hydrogen density and $c$ is the speed of light.

\begin{figure*}
  \vbox{\psfig{figure=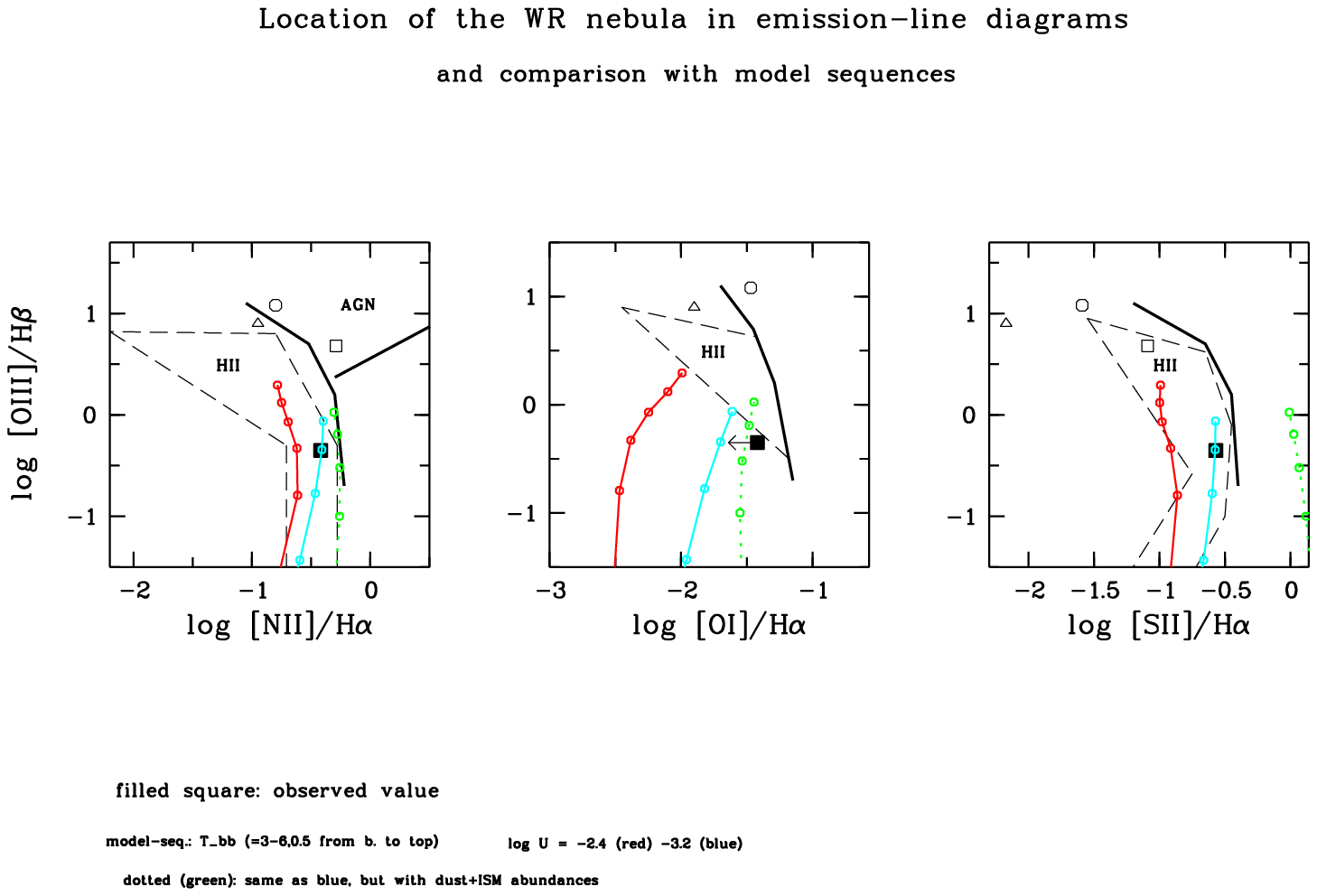,width=16.8cm,%
         bbllx=1.6cm,bblly=4.6cm,bburx=16.6cm,bbury=9.5cm,clip=}}\par
  \caption{Location of our WR nebula (filled square) in diagnostic diagrams, 
   as compared to a  few others (open symbols). The region enclosed in the 
   dashed lines marks typical
   stellar ionization as observed in HII regions and HII galaxies
   (note, that the left borderline just reflects current observation
   limits, since weak lines might have escaped detection).
   The thick line is the dividing line (Osterbrock 1989) as compared
   to other types of ionization (like a hard continuum in AGN).
   Our WR-nebula is located well within the region populated by HII regions.
   For comparison, we plot some PNs surrounding WRs using the sample of
   Pena \etal\ 1997 (note: biased to low values of [OIII] since in
   several objects of their sample the [OIII]-line was saturated, thus
   not plotted.)
%
   Results from photoionization models are plotted in these diagrams as well: 
   The lines correspond to several model sequences. 
   Along a line, T$_{\rm bb}$ varies between 30.000 and 60.000
   K in steps of 5000 K from bottom to top. 
   Each line corresponds to a fixed ionization parameter
   (log U = --2.4 (blue), log U = --3.2 (red)). The dotted (green) line 
   corresponds to a sequence were dust was mixed with the gas and the 
   metal abundances were depleted (log U = --2.4 as for the `blue' model).
   The model with log U=--3.2 and T$_{\rm bb}$=55000 K best matches 
   all observed line ratios (Table \ref{cloud}). 
    }
  \label{diagall}
\end{figure*}

A main observational feature is the relative weakness of [OIII]/H$\beta$
(though well within what is observed in `normal HII regions';
cf. Fig. \ref{diagall}). 
Whereas close distances and high temperatures ($\gax 4.5 \times 10^4$ K)
significantly overpredict the strength of \ion{[O}{III]} (cf. Fig. 8),
it is of the order of the observed value for larger distances of
the illuminated gas (i.e., lower ionization parameters).  
Further, we note that the quite substantial change in strengths of some 
emission lines, particularly \ion{[O}{III]}, over a distance 
$\sim$1 pc to $\sim$40pc
serves as argument that we are only seeing a shell of gas,
as is also suggested by the visual impression.

We further find that low-ionization lines are underpredicted
for small distances (less than a few 10$^{19}$ cm), i.e., large ionization 
parameters, consistent with general expectations (e.g. Stasinska 1982, 
Komossa \& Schulz 1997). 

\begin{table}[ht]
\caption{Results of the photoionization calculations carried out
    with the code {\em Cloudy}. The first column gives observed values of 
    emission lines (all relative to H$\beta$; Ib representatively chosen and 
    [SII] added from region III), the next columns give the model results. 
    Model (i) was 
    calculated for $T_{\rm bb} = 55\,000$K, $\log n=1.7$ and $\log U=-3.2$; 
    model (ii) for the same $T_{\rm bb}$, $\log n = 1.4$ and $\log U = -3.3$ 
    (see text for details). $T_{\rm e}$ is the gas temperature at the 
    illuminated phase of the shell in K. We note that an even better match 
    of individual lines could be obtained by further fine-tuning $r$ or 
    $T_{\rm bb}$. Given the uncertainties, particularly the approximation
    in representing the WR-star atmosphere by a blackbody, this was not 
    attempted.}
  \vspace{-0.2cm}
  \begin{tabular}{cccc}
      \noalign{\smallskip}
      \hline
      \noalign{\smallskip}
   line ratio & region Ib & model (i)   & model (ii)$^{(1)}$ \\
 \noalign{\smallskip}
 \hline
 \noalign{\smallskip}
  H$\alpha$/H$\beta$    &   2.60    &   2.66      &   2.65    \\
  ~[OIII 4363]/H$\beta$ &   0.02    &  $<10^{-3}$  &  $<10^{-3}$ \\
  HeI  4471/H$\beta$    &  $<$0.04  &   0.05      &   0.05     \\ 
  ~[FeIII 4658]/H$\beta$    &  $<$0.03  &   0.03      &   0.02   \\
  HeII 4686/H$\beta$    & $<$0.017  &  0.002     &   0.002  \\
  ~[OIII 5007]/H$\beta$ & 0.35--0.50 &  0.45      &   0.30   \\
  ~[NI 5200]/H$\beta$  &  $<$0.02     &   0.01      &   0.01   \\
  HeI 5875/H$\beta$     &    0.15  &   0.15      &   0.15   \\
  ~[OI 6300]/H$\beta$   &    0.08  &   0.05      &   0.05   \\
  ~[NII 6584]/H$\beta$  &    1.05  &   1.04      &   1.00   \\
  ~[SII 6716]/H$\beta$  &    0.45   &   0.41      &   0.43   \\
  \noalign{$\dashdot$}
  $\log U$            &           &   -3.2      &  -3.3    \\  
  T$_e$               &           &  5.8$\times$10$^3$  & 5.8$\times$10$^3$  \\
      \noalign{\smallskip}
      \hline
    \end{tabular}

\noindent{\small
    $^{(1)}$ $\log L_{\rm ion}=39.39$, $\log L_{\rm bol}=39.59$; 
             $\log L_{\rm H\beta}=37.52$. }
   \label{cloud}
   \end{table}

In order to simultaneously match the observed strengths of low- and 
high-ionization lines we require  
a temperature of $T \simeq 5.5 \times 10^{4} - 6 \times 10^{4}$ K 
(see Table \ref{cloud}) and an ionization parameter 
of $\log U \simeq -3.2$ corresponding to 
a distance of $\log r = 19.9$ cm  (cf. model (i) of Table \ref{cloud}) for 
an ionizing (beyond the Lyman limit) luminosity of $\log L = 39.39$ erg/s and 
$T = 5.5 \times 10^4$ K; or a slightly larger distance if $n$ is lowered
(cf. model (ii) of Table \ref{cloud}).  
We also carefully checked that no other, non-observed emission lines
are overpredicted by the model.

This model (we used model (i) in the following; see Table \ref{cloud}) then 
predicts (for a shell of 4$\pi$ covering) 
an H$\beta$-luminosity of L$_{\rm H\beta}$ = 3$\times$10$^{37}$ erg/s  which
overpredicts the observed value, i.e. implies a filling factor
of less than unity, of the order of 0.1.

Given that both, \ion{[N}{II]} and \ion{[S}{II]} are correctly predicted by 
this model, we do not find evidence for a deviation of the Nitrogen 
abundance from the solar value.

Re-calculating the best-fit model, now including Galac\-tic-ISM-like dust 
(the species graphite and silicate)  but non-depleted (i.e., still solar)
abundances, has only weak influence on the resulting emission line spectrum 
(dust slightly contributes to the heating, but not much, in the present case).
However, if additionally the abundances are set to the ISM value (a mean of
Cowie \& Songaila 1986, as included in {\em Cloudy}) the effect is quite 
strong. Due to the depletion of important coolants, several line-ratios 
increase in intensity (cf. Fig. \ref{diagall}). The \ion{[S}{II]} line 
becomes particularly strong, which can be partly traced back to the 
overabundance of sulphur, which happens to be twice
the solar value in the employed set of ISM metal abundances. 
Under the assumptions made, i.e. dust properties and ISM metal abundances, 
the observed emission
lines are better matched for a dust-free environment (or selectively
depleted S abundance).

A mean nebular temperature of $T \simeq 5 \times 10^3$ K is derived for
the best-fit (dust-free) model. 


The total mass of the nebula amounts to $\sim 3\times 10^2$ \msun\ assuming 
a covering factor of 0.1 (see above).

\section{Discussion and Conclusion}

With the parameters derived above for the luminosity of the WC6 star and
the size, density and emission line strength of the nebula it seems
reasonable to assume that both are physically connected.
The geometrical appearance with the nebula being centered around the WR star
supports this conclusion. The nebula itself seems to form a shell and
based on the kinematic age and mass is thought to be interstellar matter
swept up by the wind of the WR star MLA 1159.

An important question to be addressed is on the mechanism which produces
the nebular emission. The lack of filamentary \ion{[O}{III]} structures 
together with a diffuse \ion{[O}{III]} emission in seemingly shell geometry
and the identical extent
of the nebula at different emission lines 
suggests that the nebula
is not shock excited, but rather photoionized.
Also, the emission line width and the low density which is similar 
to that in typical HII regions suggest photoionization as the main
origin. Finally, the satisfactory fits obtained with the code {\em Cloudy}
to explain the relative line strengths implies that photoionization
is a reasonable assumption.

Our successful model is characterized by an ionization parameter 
$\log U \simeq -3.2$ and it is interesting to note that previous 
investigations of Wolf-Rayet nebulae (and giant HII regions) found values
for the ionization parameter close to $-3$ as well, with a rather narrow
range in $U$ (e.g., Esteban \etal\ 1993, Bresolin \etal\ 1999)
which led to the suggestion that the stellar winds cause a close scaling 
of ionizing flux and shell gas density (Shields 1986).

\begin{figure}
  \resizebox{8.8cm}{!}{\includegraphics{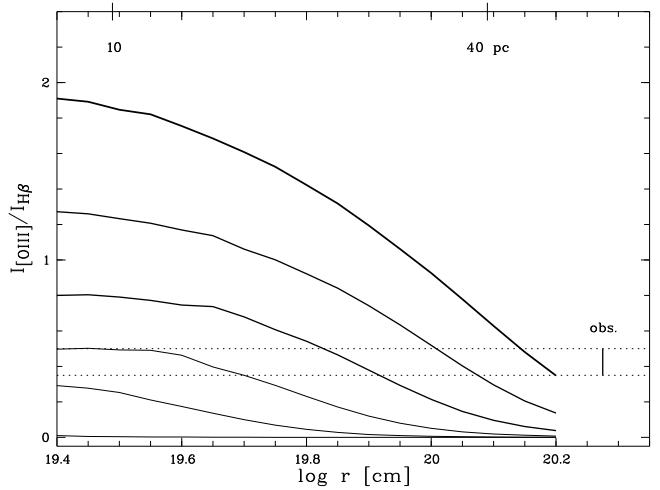}}
  \caption{Run of the [OIII]/H$\beta$ emission line ratio in dependence
     of distance $r$ (in cm) from the WR star and blackbody temperature ( =
     3, 4, 4.5, 5, 5.5, 6 $\times$ 10$^4$ K from bottom to top)
     for $\log n = 1.4$. The observed range for the emission line ratio 
     (corresponding to different slit positions) is bracketed by the dotted 
     horizontal lines.  
  }
  \label{o3run}
\end{figure}

We finally note that photoionization modelling employing blackbody continua 
for the
ionizing flux instead of real WR star atmospheres are expected to derive 
lower temperatures. This is due to the fact that the (He- and) C-rich
stellar atmosphere absorbs most of the extreme UV ionizing photons
($\lambda \lax$ 228 \AA), thus emitting many fewer UV photons than a 
blackbody of the same temperature (Koesterke \& Hamann 1997).

Because of the original goal of identifying a supersoft X-ray source we mention
here that the X-ray source \s13\ is very likely not related to either the
WC6 star or the nebula. If \s13\ were within the nebula, it would be the
dominant ionization source and produce very strong \ion{[O}{III]} emission,
i.e. \ion{[O}{III]}/H$\beta$$\approx$10--20 which is not observed.
With a bolometric (blackbody approximation) luminosity of \s13\ of
the order of 10$^{37}$ \ergsec\ the X-ray emission is also very unlikely
produced by the WC6 star which typically have X-ray luminosities around
a few times 10$^{32}$ \ergsec. Deeper imaging is necessary to identify the 
counterpart of \s13\ which we now suspect North-East of the nebula BA 1-642.

\begin{acknowledgements}
We thank G.J. Ferland for providing the code {\em Cloudy} and the referee,
A.F.J. Moffat, for a careful reading of the manuscript.
J.G. and St.K. are supported by the German Bundesmi\-ni\-sterium f\"ur Bildung,
Wissenschaft, Forschung und Technologie
(BMBF/DLR) under contract Nos. FKZ 50 QQ 9602\,3 and 50 OR 9306\,5.
GT acknowledges CONACYT grant 25454-A.
The \ros\, project is supported by BMBF/DLR and the Max-Planck-Society.
This research has made use of the Simbad database, operated at CDS, Strasbourg.
The Digitized Sky Survey was produced at the Space
Telescope Science Institute under US Government grant NAG W-2166.
\end{acknowledgements}

\end{document}